\documentclass[a4paper,11pt,twoside]{article}
\usepackage[a4paper,left=2.2cm,right=2.2cm,top=2.3cm,bottom=2.6cm]{geometry}
\usepackage[english]{babel}
\usepackage{setspace}
\usepackage{mathrsfs,hyperref, algorithm, algorithmic}
\usepackage{harvard}
\usepackage[]{amsmath}
\usepackage{amsthm}
\usepackage{fix-cm}
\usepackage[]{amssymb}
\usepackage[]{latexsym}
\usepackage[latin1]{inputenc}
\usepackage[right]{eurosym}
\usepackage[T1]{fontenc}
\usepackage[]{graphicx}
\usepackage{wrapfig}
\usepackage[]{epsfig}
\usepackage{fancyhdr}
\usepackage{bbm}
\usepackage{pstricks}
\usepackage{multirow}
\usepackage[font=small]{caption}
\usepackage{subcaption}

\makeatletter

\renewcommand{\p@enumi}{theenumi-}
\renewcommand{\@fnsymbol}[1]{\@arabic{#1}}

\newcommand{\ci}{\citeasnoun}

\makeatother
\newtheorem{lemma}{Lemma}[section]

\newtheorem{example1}[lemma]{Example}
\newtheorem{ex1}[lemma]{Example}
\newtheorem{rem1}[lemma]{Remark}

\newtheorem{alg1}[lemma]{Algorithm}
\newtheorem{me1}[lemma]{Mechanism}

\usepackage{color}

\numberwithin{equation}{section}
\numberwithin{figure}{section}
\numberwithin{table}{section}

\begin{document}

\title{It's a Trap: Emperor Palpatine's Poison Pill}
\author{ Zachary Feinstein\footnote{Zachary Feinstein,  ESE, Washington University, St. Louis, MO 63130, USA, {\tt zfeinstein@ese.wustl.edu}.}\\[0.7ex] \textit{Washington University in St. Louis}}
\date{\today}
\maketitle

\addtocounter{footnote}{1}
\begin{abstract}
In this paper we study the financial repercussions of the destruction of two fully armed and operational moon-sized battle stations (``Death Stars'') in a 4-year period and the dissolution of the galactic government in \emph{Star Wars}.  The emphasis of this work is to calibrate and simulate a model of the banking and financial systems within the galaxy.  Along these lines, we measure the level of systemic risk that may have been generated by the death of Emperor Palpatine and the destruction of the second Death Star.  We conclude by finding the economic resources the Rebel Alliance would need to have in reserve in order to prevent a financial crisis from gripping the galaxy through an optimally allocated banking bailout.\footnote{No Bothans died to bring us this information.}
\end{abstract}\vspace{0.2cm}

\section{Introduction}\label{Sec:intro}

Economics and finance, much like the Force as explained by Jedi Master Obi-Wan Kenobi, is ``created by all living things.  It surrounds us and penetrates us; it binds the galaxy together'' (\ci{Ep4}).  Emperor Palpatine, as a ruler of multiple decades, understands this concept as well as any other successful politician must.  Further, as he demonstrated by successfully playing both sides during the Clone Wars, Emperor Palpatine must assuredly be able to create contingency plans.  As such, Emperor Palpatine would have had a plan in case he were going to be defeated.  This paper focuses on the logic that Emperor Palpatine would have considered in order to avoid the outcome of the Battle of Endor in 4ABY (\emph{Star Wars: Episode VI -- Return of the Jedi}), i.e., in which Emperor Palpatine and Sith Lord Darth Vader have been killed and a massive government project (the second Death Star) has been destroyed via mutually assured destruction with the Rebel Alliance.

In fact, Darth Vader recognized that moon-sized space stations were not as all-powerful as the force-insensitive believed; he quipped in response to the first Death Star,
``Don't be too proud of this technological terror you've constructed. The ability to destroy a planet is insignificant next to the power of the Force'' (\ci{Ep4}).
Lord Vader's logic can be taken further, as we conjecture Emperor Palpatine considered and we will show, to prove that the true power of the technological terror is from the power of economic terror unleashed in the case of its failing; i.e., in case of the Rebel Alliance succeeding, an automatic financial attack on the entire Galaxy would be realized.

Emperor Palpatine misjudged one point in his strategy of mutually assured destruction: the Rebel Alliance, while trusting in the power of the Force, never placed importance on long-term planning.  As General Han Solo of the Rebel Alliance, well-known for shooting first,\footnote{Despite the claims of some revisionist historians.} once famously said, ``Never tell me the odds'' (\ci{Ep5}).  As a high-ranking official, this viewpoint would have been widespread within the Rebel Alliance.  Since mutually assured destruction relies on both parties recognizing it as such, and because the Rebels do not recognize it as such due to their own poor long-term thinking, Emperor Palpatine's economic warfare strategy is an ineffective deterrent.

The analysis of this paper is as follows.  First, we calibrate a model of the economic health and financial system in the galaxy during the Imperial Period.  Second, we consider a distribution of economic losses in the Imperial banking system following the events of the Battle of Endor to find the risks that the Rebel Alliance has created.  Finally, we use these results to deduce the size and composition of a bailout necessary that the new rulers of the galaxy, i.e., the Rebel Alliance, would need to allocate to stave off a galaxy-wide financial crisis and economic depression.

\section{Modeling the Galactic Economy}

As reported in \ci{deathstarcost}, and quoted by the United States government (\ci{whitehouse}), the cost for the steel\footnote{Given technological improvements we assume the cost of durasteel is comparable to that of steel on Earth.} of the DS-1 Orbital Battle Station [DS1] (the first Death Star) would come out to \$852 QUADRILLION priced in Earth year 2012.\footnote{Assuming a diameter of 140km.  In \ci{cross-sections} the diameter of DS1 is reported as 160km, at this size the price of steel would be the significantly higher \$1.272 QUINTILLION in 2012 dollars.}
\captionsetup[wrapfigure]{skip=3pt}
\begin{wrapfigure}{R}{0.5\textwidth}
\centering
\includegraphics[width=0.49\textwidth]{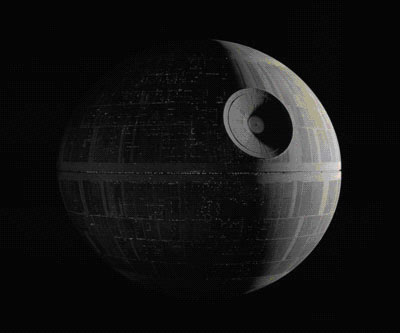}
\caption{Not a moon!}
\label{Fig:DeathStar1}
\end{wrapfigure}
We will assume that the costs for all other components are in line with that of the USS Gerald Ford, the most recently completed aircraft carrier for the United States.  For simplification purposes, we will assume the full \$17.5 BILLION cost for the USS Gerald Ford (cf.\ \ci{GeraldFord}) at approximately 100,000 tonnes (cf.\ \ci{GeraldFordDisplacement}) has the same density as the HMS Illustrious.\footnote{The density of the HMS Illustrious was used for the cost analysis presented in \ci{deathstarcost}.}  Expanding upon the logic from \ci{deathstarcost}, we assume the ratio of steel cost to total cost is the same as with the USS Gerald Ford.  Therefore we get an estimated \$193 QUINTILLION cost for the Death Star (including R\&D).  With the same analysis, the completed second Death Star [DS2] with its reported 900km diameter (cf.\ \ci{insideTrilogy}) would be \$226 QUINTILLION in steel alone.  However, we will assume the components beyond basic infrastructure for DS2 were comparable to DS1 for a total cost of \$419 QUINTILLION.  Note that this is a lower bound, though there is more weaponry we ignore any new research and development costs.\footnote{Using the same cost formula as DS1, the cost of DS2 would be \$51.4 SEXTILLION.}  
However, this figure is disputed by \ci{starwarsWH} which quotes Admiral Conan Motti of the Imperial Starfleet as stating,
``The costs of construction they cited were ridiculously overestimated, though I suppose we must keep in mind that this miniscule [sic] planet does not have our massive means of production.''
Admiral Motti has the correct idea, as the Galactic economy is surely larger than that of the United States of America in 2012.  We will use these numbers however to generate an economic model for the size of the Galactic economy.

For calibration purposes we will use the \$193 QUINTILLION figure, but normalize it to Gross Galactic Product [GGP].  As a massive military research and development project, we will make the analogy between the construction of DS1 and the construction of the first atomic bomb.  That is, assume the construction for the first Death Star had a similar cost profile in gross terms to the Manhattan Project.  From \ci{manhattanProject} we get an estimate for \$2.2 BILLION for the cumulative project in 1945 dollars.  The cost figure needs to be put into perspective of the time in which it was applied.  Table~\ref{Table:USGDP} provides the expenditures for the Manhattan Project from and the Gross Domestic Product [GDP] of the United States from 1942-1946 in 1945 dollars (see \ci{BEAGDP}).
\begin{table}[t]
\centering
\begin{tabular}{|r|*{5}{c}|c|}
\hline
Year & 1942 & 1943 & 1944 & 1945 & 1946 & \textbf{Total}\\ \hline
Expenditures (MILLION \$) & 16.1 & 344.6 & 939.4 & 610.3 & 281.0 & 2,191.4 \\
GDP (BILLION \$) & 182.5 & 213.2 & 230.3 & 228.2 & 202.4 & 1,056.6 \\ \hline
Percent GDP & 0.01\% & 0.16\% & 0.41\% & 0.27\% & 0.14\% & 0.21\% \\ \hline
\end{tabular}
\caption{Manhattan Project Expenditures and US GDP from 1942-1946 in 1945 Dollars.}
\label{Table:USGDP}
\end{table}  
The first Death Star took 20 years of work to complete.  Assuming a similar cost profile over time as was used for the construction of the first atomic bomb we estimate the size of investment as approximately 0.21\% of the GGP over the 20 years.\footnote{If we exclude 1942, due to its relatively low expenditures, we arrive at nearly 0.25\%.  While not an insignificant change, we will assume the first years of the construction of DS1 would have been relatively low-cost as well.}  At \$193 QUINTILLION, that puts the total GGP over the 20 year period at approximately \$92 SEXTILLION, or an average of approximately \$4.6 SEXTILLION per year.\footnote{To put this number into perspective: Gross World Product for Earth in 2014 was approximately \$78 TRILLION (see \ci{WorldBankGDP}) and \$16,100 per capita (see \ci{CIAGDP}); whereas GGP/planet (over 1.75 million full member worlds of the Empire (cf.\ \ci{EssentialAtlas})) would be \$2.6 QUADRILLION and \$46,000 per capita (over 100 quadrillion inhabitants (cf.\ \ci{EssentialAtlas})).  Though the numbers for GGP/planet and GGP/capita may seem small in relation to the technology available, recall that until the industrial revolution income/capita was stable (cf.\ \ci{IndustrialRevolution}).  In a galaxy in which technology has not progressed for 25,000 years (\ci{techProgress}) it is not unreasonable to assume GGP grew at a much slower pace, if at all, during the Galactic Republic than it has on Earth since the Industrial Revolution.} 

The GGP numbers are not important for our purposes in and of themselves.  Instead we wish to use these numbers in order to determine the size of the banking and financial sector of the Galactic Empire circa 4ABY at the destruction of DS2 and the death of Emperor Palpatine during the Battle of Endor. 

First we will consider the size of the Intergalactic Banking Clan [IGBC] which is known to be ``too big to fail'' as it was nationalized following the Clone Wars to avoid a default on its debt (cf.\ \ci{wookieepediaIGBC}).  In order to determine the size of IGBC we compare it to activities of the Federal Reserve in the United States during periods of intervention.  In particular, as reported in \ci{FedBalanceSheet}, during the Great Depression and the Second World War the Federal Reserve balance sheet grew to have assets in excess of 20\% of nominal GDP.  After nationalizing IGBC it acted as a creditor of the Empire's currency, so we will assume that following the Clone Wars it held a similar 20\% of GGP.

For the remainder of the banking and financial sector of the Galactic Empire we assume, as per \ci{wookieepediaBank}, ``there were many banks in the galaxy.''  Given 1.75 million full member worlds of the Galactic Empire (cf.\ \ci{EssentialAtlas}), we will make the assumption that for every 100 worlds there is a ``big'' bank.  Further, 1 of every 100 of these banks is of a ``massive'' size, but not ``too big to fail.''\footnote{These might include any number of the banks referred to in \ci{listBanks}.}  This produces 17,501 banks in the galaxy for our model.  We will assume all other banks are of smaller size and not of systemic importance.  As a simplification we will assume that each subgroup of banks (IGBC, ``massive'', and ``big'') encompasses a full 1/3 of the entire banking sector.  In total this implies that the banking sector has assets of 60\% of GGP.\footnote{Compared to US and Europe this is a low number (cf.\ \ci{helgi}), though it is notably higher than the percentage of Earth GDP.  Not surprising as banking is not the most important sector of the Galactic economy.}$^,$\footnote{With fractional reserve banking and deregulated banks (we refer to \ci{deregulationShow}) this could mean very high leverage ratios, not necessarily that the banks have deposits worth 60\% of the economy.}

\section{Modeling the Financial Crisis and a Desired Bailout Allocation}

Now that we have the economic situation of the Galactic Empire, we wish to model the crisis following the destruction of DS2 and the death of Emperor Palpatine.  Firstly we will assume that already 50\% of the cost of DS1 has been paid off, but none of the cost of DS2 has been paid.\footnote{A very aggressive repayment schedule for DS1 debt, thus providing a lower bound on the size of the resultant crisis.  Compare with the UK having not fully payed off its debt from WW1 by 2014 (cf.\ \ci{UKWW1}).}  Additionally there was no other Galactic debt which the Empire had outstanding.  This is assuming that Emperor Palpatine was fiscally conservative and kept the overall debt load under control.  Further, we assume that the interest rate the Galactic Empire must pay on its debt is 0\%.\footnote{This provides us a lower bound on the size of losses incurred, so any other value that we may choose would only cause the Galactic Economy to fare worse.}  Therefore, with the disintegration of the Galactic government and the destruction of both DS1 and DS2, the losses from governmental bond defaults measure \$515.5 QUINTILLION.\footnote{This is a loss of approximately 3\% of GGP under the rosiest of scenarios (10\% annual GGP growth) and greater than 10\% of GGP assuming 0\% average growth.}

For our analysis we will assume a relatively high-growth environment given the political situation, under the assumption the Empire was able to kickstart the Galactic economy.  We can make this assumption as it provides a lower bound to the risks, i.e., a lower growth environment would have larger economic losses as a fraction of the economy. Let us assume an average GGP growth of 2\%; GGP in 4ABY would have a value of approximately \$6,090 QUINTILLION.  This is a considerably high-growth rate for the Galactic Empire.  We come to this conclusion because of the nonexistent technological progress for 25 millennia (\ci{techProgress}) but also the political instability evidenced by the Rebel Alliance's existence.\footnote{As stated in \ci{legitimacyBook}, ``A large literature in political economy, for instance, finds that short- or medium-term fluctuations in economic growth can have a profound effect on state legitimacy.''} Further, we assume away the possible economic repercussions from the Battle of Yavin and the destruction of DS1 (\emph{Star Wars: Episode IV -- A New Hope}).

Following the destruction of DS2, we will assume there are random losses to asset values due to the ``terrorist'' attack.  Following the 9/11 terrorist attacks, the Dow Jones Industrial Average dropped 14.2\% and the S\&P500 was down 11.6\% in the week following the market reopening (cf.\ \ci{investopedia911}).  Additionally, a fake Tweet from the Associated Press about a terror attack on the White House on April 23, 2013 caused a nearly 1\% drop in the market in a 4-minute window before the information was corrected (cf.\ \ci{marketwatchTweet}).  Given these data points, and the ``success'' of the Battle of Endor for the Rebel Alliance, we will assume that the market drops by 20\% (on average) causing losses before considering the default on government debt.\footnote{Each firm's valuation follows a $beta(1,4)$ percent loss with 25\% correlations via a Gaussian copula.}  Due to fractional reserve banking we will assume that assets are 4-times total deposits.   This is a low (conservative) requirement,\footnote{Typically 3\% or 10\% reserve requirements are imposed in the United States in 2015 (\ci{reserveReq}).} but this is a reasonable estimate as banking is a smaller part of the Galactic economy.

Finally, while we assume there was deposit insurance for all the banks under the rule of the Galactic Empire (i.e., depositors are able to retain their capital in case of a bank failure), with the fall of the government this can no longer be assumed to exist.  Thus if a firm were to have a shortfall on its obligations, all deposits with that bank are lost as well.

\begin{wrapfigure}{R}{0.5\textwidth}
\centering
\includegraphics[width=0.49\textwidth,trim={1.4cm 0 1.4cm 0},clip]{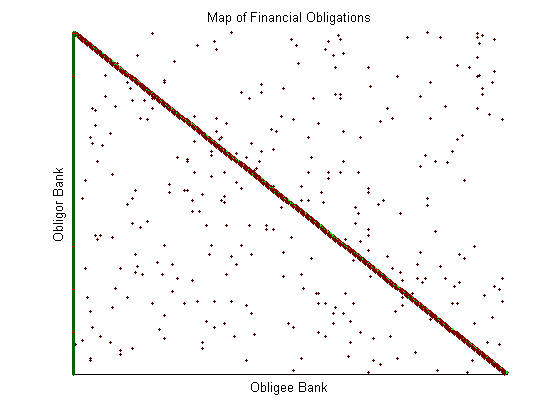}
\caption{A graphical representation of the connections in the banking sector.}
\label{Fig:financial_network}
\end{wrapfigure}

To simulate the banking sector we assume that the financial system has connections as displayed in Figure~\ref{Fig:financial_network} and the realized payments follow the model of \ci{EN01}.\footnote{IGBC owes \$2500 QUINTILLION to the economy outside of the ``big'' and ``massive'' firms considered here, and we will assume that none of the remaining 17,500 ``big'' and ``massive'' firms have any obligations outside of the financial system.  Further, IGBC owns 2/3 of all government debt from the Death Star (as the central bank and primary creditor of the Galactic Empire).  The remaining 1/3 is distributed evenly amongst the 175 ``massive'' banks.  When a ``massive'' bank has a liability they owe \$3 QUINTILLION, \$333 QUADRILLION, and \$500 QUADRILLION respectively to IGBC, ``massive'', and ``big'' banks.  Similarly the ``big'' banks have liabilities of size \$100 QUADRILLION, \$470 QUADRILLION, and \$2 QUADRILLION respectively to IGBC, ``massive'', and ``big'' banks.}  In particular, we are interested in the size and composition of the bailout that the Rebel Alliance would need to provide in order to keep the Galactic economy from crashing.  In order to accomplish this, we use the systemic risk measures proposed in \ci{feinstein2014measures}.  However, to complicate matters we assume that the IGBC will never directly receive any bailout funds.  This is because the IGBC is too big to fail, but without a central government this also implies that it is too big to save.\footnote{The politics of bailing out the IGBC could also be difficult due to its close relationship with the Galactic Empire.}

\begin{wrapfigure}[11]{l}{0.5\textwidth}
\centering
\includegraphics[width=0.49\textwidth,trim={0.64cm 0.2cm 1.25cm 0.2cm},clip]{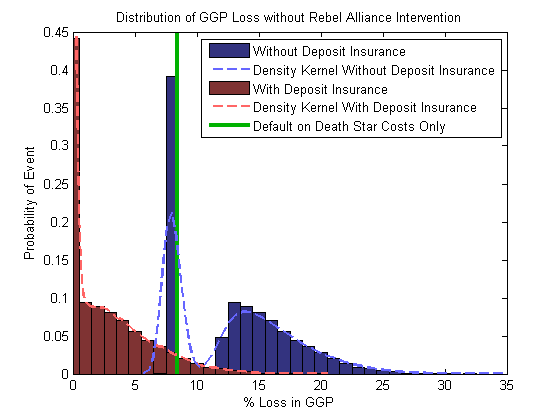}
\caption{Distribution of the losses caused by the destruction of DS2.\\~\\}
\label{Fig:percggp}
\end{wrapfigure}

In the case that the Galactic Empire wins the Battle of Endor, thus having no default on government bonds, no drop in asset values, and having deposit insurance, there is 0\% loss of GGP from the financial system though some banks are unable to pay their debts in full; while it is non-negligible to invoke deposit insurance, this is not a destabilizing force as evidenced by the 0\% drawdown in GGP to the ``real economy''.  However, once DS2 is blown up and Emperor Palpatine is killed, the GGP drawdowns are depicted in Figure~\ref{Fig:percggp}.\footnote{From 1929-1933 the United States real GDP fell 26.7\%, distributed across all industries (\ci{depressionStats}). In comparison, the losses presented here occur in the immediate aftermath in only the financial sector.}  In this image we plot the empirical 
distribution of the economic losses in our system as a histogram under the setting without deposit insurance (blue bars) and again with deposit insurance (maroon bars).  Additionally, the green line in the graph shows the losses caused by the government default on debt from DS1 and DS2 only, without any financial system included.  The dotted lines show a best approximation of a continuous probability density function fit to the simulated data to smooth the shape.  What is most noticeable is in nearly 40\% of our simulations the financial system absorbs much of the losses and generally benefits the real economy by being smaller than the \$515.5 QUINTILLION in government bond defaults.  However, the economic outlook is still bleak in these cases, with losses greater than 7\% of GGP, it is however an improvement over the case in which the citizens of the Galaxy directly financed 
the building of DS1 and DS2 through bond purchases.  In the other 60\% of the simulations, the network effects amplify the losses to a significant margin.  In fact, we find in the cases of amplified losses, greater than 99\% of all systemically important banks default on a portion of their obligations.  Though with deposit insurance we notice that the losses incurred are far reduced, there are still significant tail risks.  In expectation, with deposit insurance, there is still a nearly 2.5\% drop in GGP.\footnote{In comparison, during the 4th quarter of 2008, GDP in the United States dropped by roughly 1.6\% (\ci{recessionStats}).}  In fact, as deposit insurance is not free, in order to attain these results requires an investment of roughly 8\% of GGP in the ``good'' (approximately) 40\% of cases and 12\% of GGP in the remaining 60\% of situations. For deposit insurance to be credible, the 12\% of GGP would need to be held in reserve by the Rebel Alliance.  Otherwise, when the situation arose where depositors demand their funds from the banks and subsequently from the deposit insurance, the Rebel Alliance would be shown to be making empty promises.  As mentioned previously, even with this expensive deposit insurance the losses to the Galactic economy are still extremely high (an expected 2.5\% drop in GGP) so an additional banking bailout would be necessary to rescue the economy.

\begin{table}
\centering
\begin{tabular}{|r|*{4}{c|}}
\hline
& \multicolumn{1}{|p{3.5cm}|}{\centering ``Massive'' \\ (QUADRILLIONS)} & \multicolumn{1}{|p{3.5cm}|}{\centering ``Big'' \\ (QUADRILLIONS)} & \multicolumn{1}{|p{3.15cm}|}{\centering Total \\ (QUINTILLIONS)} & \multicolumn{1}{|p{2cm}|}{\centering Total \\ (\% GGP)} \\ \hline
Expectation & \$2,813 & \$26 & \$938 & 15.4\% \\ \hline
Value-at-Risk & \$3,227 & \$31 & \$1,110 & 18.2\% \\ \hline
Average VaR & \$3,882 & \$37 & \$1,312 & 21.5\% \\ \hline
\end{tabular}
\caption{Minimal size for a bailout under the three criteria considered.  All values rounded to the nearest whole number.}
\label{Table:orthant}
\end{table}

Finally we consider the size and composition of the bailout that would be necessary to rescue the financial system.  As stated before, we will assume this in the case with no deposit insurance (in fact the bailout will act as a type of insurance independent of system state, and therefore is implicitly considered).  We consider three distinct criteria that may be used to determine the size and composition of the necessary bailout:
\begin{enumerate}
\item bailout compositions so that the expected losses to the Galactic economy are no worse than 1\% of GGP;
\item bailout compositions so that the losses to the Galactic economy are worse than 1\% of GGP with probability less than 10\% (called the Value-at-Risk);
\item bailout compositions so that the expected losses to the Galactic economy in the worst 10\% of cases are no worse than 1\% of GGP (called the Average Value-at-Risk [Average VaR]).
\end{enumerate}

\begin{wrapfigure}{l}{0.5\textwidth}
\centering
\includegraphics[width=0.49\textwidth,trim={0.4cm 0.2cm 1.35cm 0.2cm},clip]{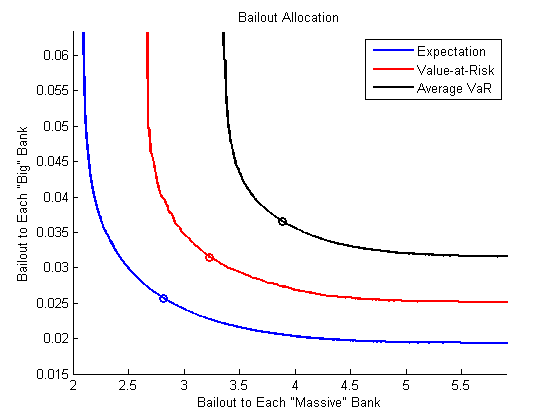}
\caption{Necessary bailouts in QUINTILLIONS. The minimal total bailout packages are circled.}
\label{Fig:riskmsr}
\end{wrapfigure}
\noindent Figure~\ref{Fig:riskmsr} displays the composition of the minimal bailouts in the amount distributed to each ``massive'' firm (x-axis) and each ``big'' firm (y-axis) so that the three criteria are satisfied.  Notice that as we increase the size of the bailout to the ``big'' firms we can decrease the size to the ``massive'' firms, and vice versa, but with diminishing returns.  The circled values show the minimal total bailout package under each criteria.  Table~\ref{Table:orthant} provides the allocations in each scenario and the total size of the bailout (both in nominal and in relative terms) by utilizing the multipliers of 175 ``massive'' banks and 17,325 ``big'' banks in our model.  In order to obtain an expected loss in GGP of at most 1\% would require greater than 15\% of GGP; to achieve at most 1\% expected losses in only the worst 10\% of cases would require greater than 20\% of GGP.  These are values that the Rebel Alliance would need to have prepared in order to avoid the catastrophic financial collapse evidenced in Figure~\ref{Fig:percggp}.  Given that the Rebel Alliance is the ``scrappy underdog,'' without the resources to build multiple moon-sized space stations, this is a sum they do not have.

\section{Conclusion}
We modeled the state of the economy of the Galactic Empire prior to the destruction of the two moon-sized battle stations and the fall of the Imperial government.  This allowed us to calibrate a financial network of the systemically important institutions, thus providing a picture of the economic repercussions from the Battle of Endor.  In this case study we found that the Rebel Alliance would need to prepare a bailout of at least 15\%, and likely at least 20\%, of GGP in order to mitigate the systemic risks and the sudden and catastrophic economic collapse.  Without such funds at the ready, it likely the Galactic economy would enter an economic depression of astronomical proportions.

{\footnotesize
\bibliography{bibtex2}
\bibliographystyle{jmr}}

\end{document}